\documentclass[aps,prd,reprint,nofootinbib,twocolumn,superscriptaddress,preprintnumbers]{revtex4-2}

\usepackage{amsthm}
\usepackage{amsmath}
\usepackage{graphicx}
\usepackage{slashed}
\usepackage{amssymb}
\usepackage[colorlinks=True, citecolor=blue, urlcolor=blue, linkcolor=blue]{hyperref}
\usepackage{multirow}
\usepackage{orcidlink}
\usepackage{qcircuit}
\usepackage{physics}
\usepackage{booktabs}
\usepackage{wasysym}
\usepackage[caption=false]{subfig}
\usepackage{ulem}
\graphicspath{{./figs/}}

\newcommand{\nn}{\nonumber}
\newcommand{\bea}{\begin{align}}
\newcommand{\eea}{\end{align}}

\begin{document}
\title{Spin Hamiltonian for Quantum Simulation of Minimally Truncated $3+1$D SU(2) Lattice Gauge Theory and Testing Eigenstate Thermalization}

\author{Ziyuan Li}
\email{ziyuanli@uw.edu}
\affiliation{InQubator for Quantum Simulation, Department of Physics, University of Washington, Seattle, Washington 98195, USA}

\author{Xiaojun Yao}
\email{xjyao@uw.edu}
\affiliation{InQubator for Quantum Simulation, Department of Physics, University of Washington, Seattle, Washington 98195, USA}

%\date{\today}
\preprint{IQuS@UW-21-137}

\begin{abstract}
We construct the spatial-gauge-singlet Hamiltonian of pure SU(2) lattice gauge theory in $3+1$ dimensions ($3+1$D) on qubits under the minimal truncation $j_{\rm max}=1/2$, building on previous work on the hyperhoneycomb lattice, and test the eigenstate thermalization hypothesis (ETH). By enforcing Gauss's law and exploiting the redundancy among gauge-invariant loop operators, we find the minimal set of independent loop variables that can be bijectively mapped onto qubits, and write the Hamiltonian as a sum of Pauli strings. An $n\times n\times n$ lattice with fixed boundary conditions requires $2(n-1)^3+3(n-1)^2$ qubits. We exactly diagonalize the Hamiltonian on a $2\times 3\times 3$ lattice and identify the coupling range of quantum ergodicity from the level statistics. At the ergodic coupling $g=0.3$, we calculate the matrix elements of local energy density operators and the entanglement entropies of all eigenstates and find them consistent with the ETH. These results show that the $3+1$D truncated SU(2) lattice gauge theory thermalizes in accordance with the ETH and provide a benchmark for future real-time quantum simulations of non-equilibrium dynamics.
\end{abstract}

\maketitle

\section{Introduction}
The rapid development of quantum computing has motivated extensive studies of the Hamiltonian lattice approach to field theories~\cite{Zohar:2015hwa,Banuls:2019bmf,Klco:2021lap,Bauer:2022hpo,Bauer:2023qgm,DiMeglio:2023nsa,Beck:2023xhh,Halimeh:2025vvp,Banuls:2013jaa,Buyens:2013yza,Kuhn:2015zqa,Pichler:2015yqa,Magnifico:2019kyj,Honda:2021aum,Dempsey:2022nys,Dempsey:2023gib,Jha:2024jan,Florio:2025hoc,Banuls:2025wiq,Abel:2025pxa,Grieninger:2025mbm,Gupta:2026tcg,Fujikura:2026tqk,Yao:2026yld,Farrell:2024fit,Zemlevskiy:2024vxt,Illa:2024kmf,Farrell:2024mgu,Hayata:2024smx,Farrell:2025nkx,Chen:2025zeh,Abel:2025zxb,Li:2025sgo,Davoudi:2025rdv,Huie:2025yzn,Chai:2025kbi,Chen:2026tvd,Andrade:2026kbz,Gonzalez:2026loy,Hayata:2026tly,Osborne:2022jxq,Ciavarella:2023mfc,Itou:2024psm,Mueller:2024mmk,Maiti:2024jwk,Ale:2024uxf,Turro:2024pxu,Spagnoli:2024mib,Budde:2024rql,Ebner:2024qtu,Yang:2025edn,DiMarcantonio:2025cmf,Illa:2025njz,NSSrivatsa:2025jhh,Balaji:2025afl,Balaji:2025yua,Ale:2025sxz,Yao:2025cxs,Cobos:2025krn,Horn:2026bfs,Orlando:2026ten,Cao:2026qky,Majcen:2026sfm,Joshi:2026hfe,Xu:2026ibi,Lewis:2026ync,Rouxinol:2026vjl,Rouxinol:2026eur,Turco:2026cte,Ale:2026hqe,Verde:2026hcg}, in the hope that it will enable simulation of real-time dynamics in the nonperturbative regime and unveil the mystery of rapid local thermalization in far-from-equilibrium scenarios such as the early stage in heavy-ion collisions.

For lattice gauge theories (LGTs) that are more directly relevant to the Standard Model, various Hamiltonian formulations have been proposed and studied, including the temporal-gauge Kogut-Susskind (KS) Hamiltonian~\cite{Kogut:1974ag} in the electric basis~\cite{Klco:2019evd,Ciavarella:2021nmj,Hayata:2021kcp,ARahman:2021ktn,ARahman:2022tkr,Muller:2023nnk,Kavaki:2024ijd,Illa:2025dou,Chen:2026hnh}, a mixed-basis approach~\cite{Grabowska:2024emw,DAndrea:2023qnr,Froland:2025bqf} based on the maximal-tree gauge~\cite{Creutz:1976ch}, the loop-string-hadron formulation~\cite{Raychowdhury:2019iki,Kadam:2022ipf,Kadam:2024ifg,Burbano:2024uvn}, the large-$N_c$ expansion~\cite{Ciavarella:2024fzw,Ciavarella:2025bsg,Modi:2026syn}, $q$-deformed algebras~\cite{Zache:2023dko,Hayata:2023puo,Hayata:2023bgh,Hayata:2024fnh,Hayata:2026rmv}, finite subgroups of SU($N_c$)~\cite{Gustafson:2023kvd,Lamm:2024jnl,Perez:2025cxl}, variationally optimized coupling-dependent bases~\cite{Haase:2020kaj,Fontana:2024rux,Miranda-Riaza:2025fus}, the orbifold formalism~\cite{Bergner:2024qjl,Halimeh:2024bth,Bergner:2026duh}, and non-temporal-gauge formulations~\cite{Li:2024ide,Yao:2025uxz,Yao:2026rya,Li:2026ppp}. Most previous numerical studies have focused on lower-dimensional theories; exceptions include studies of the phase structure of $3+1$ dimensional ($3+1$D) U(1) LGT~\cite{Magnifico:2020bqt,Kan:2021nyu} and real-time simulations of truncated SU(2) LGT in $3+1$D on a single cube~\cite{Jiang:2025ufg}, a single triamond unit cell~\cite{Kavaki:2024ijd}, and three triamond unit cells in a row~\cite{Kavaki:2025hcu}.

From the perspective of studying non-equilibrium dynamics, one would like to test the Eigenstate Thermalization Hypothesis (ETH)~\cite{PhysRevA.43.2046,PhysRevE.50.888,rigol2008thermalization}, simulate thermalization and hydrodynamization, and extract transport coefficients. The interesting physics in such studies typically emerges only when the Hilbert space and lattice volume are sufficiently large, and it certainly cannot be probed with a single unit cell in $3+1$D. This is why previous studies of SU(2) non-equilibrium dynamics have been restricted to $1+1$D and $2+1$D, including tests of the ETH~\cite{Yao:2023pht,Ebner:2023ixq,Das:2025utp}, calculations of entanglement entropy~\cite{Ebner:2024mee,Ebner:2024qtu}, simulations of thermalization~\cite{Ebner:2025pdm,Chen:2026tvd} and hydrodynamization~\cite{Turro:2025sec}, and calculations of shear viscosity~\cite{Turro:2024pxu}.

In this work, we take a first step towards simulating interesting non-equilibrium dynamics for $3+1$D SU(2) LGT. We use the hyperhoneycomb lattice on which the SU(2) LGT has been constructed in previous work~\cite{Illa:2025dou}. The advantage of the hyperhoneycomb lattice and trivalent lattices in general is that spatial gauge singlet states can be uniquely specified by the irreducible representation labels on each link. Moreover, under the minimal truncation $j_{\rm max}=1/2$ that we will adopt in our numerical calculations, the physical Hilbert space, consisting only of SU(2) singlet states that satisfy Gauss's law at every vertex, can be bijectively mapped onto qubits. The Hamiltonian in this physical Hilbert space can then be written out explicitly in terms of Pauli matrices for an arbitrary lattice size. After establishing these, we exactly diagonalize the Hamiltonian on a $2\times 3\times 3$ hyperhoneycomb lattice, which corresponds to 16 qubits under fixed boundary conditions, and analyze the level statistics to identify the range of couplings in which the system is quantum ergodic. Finally, we choose a coupling in the ergodic regime and test the ETH by computing the matrix elements of the local Hamiltonian density in the energy eigenbasis and the entanglement entropies of all eigenstates.

The paper is organized as follows. In Sec.~\ref{sec:H}, the SU(2) KS Hamiltonian on a hyperhoneycomb lattice will be reviewed and its matrix elements in the electric basis will be shown, the latter of which are new here. In Sec.~\ref{sec:H_onehalf}, we will consider the minimal truncation, map the physical Hilbert space onto qubits, and specify the lattice Hamiltonian completely in terms of Pauli matrices. Then, in Sec.~\ref{sec:results}, we will present exact diagonalization results for level statistics and ETH tests on a $2\times3\times3$ lattice. Finally, we will conclude in Sec.~\ref{sec:conclusions} with a future perspective on real-time quantum simulation.

\section{Kogut-Susskind Hamiltonian on Hyperhoneycomb Lattice}
\label{sec:H}
\begin{figure}[h]
  \centering
  \includegraphics[width=0.95\linewidth]{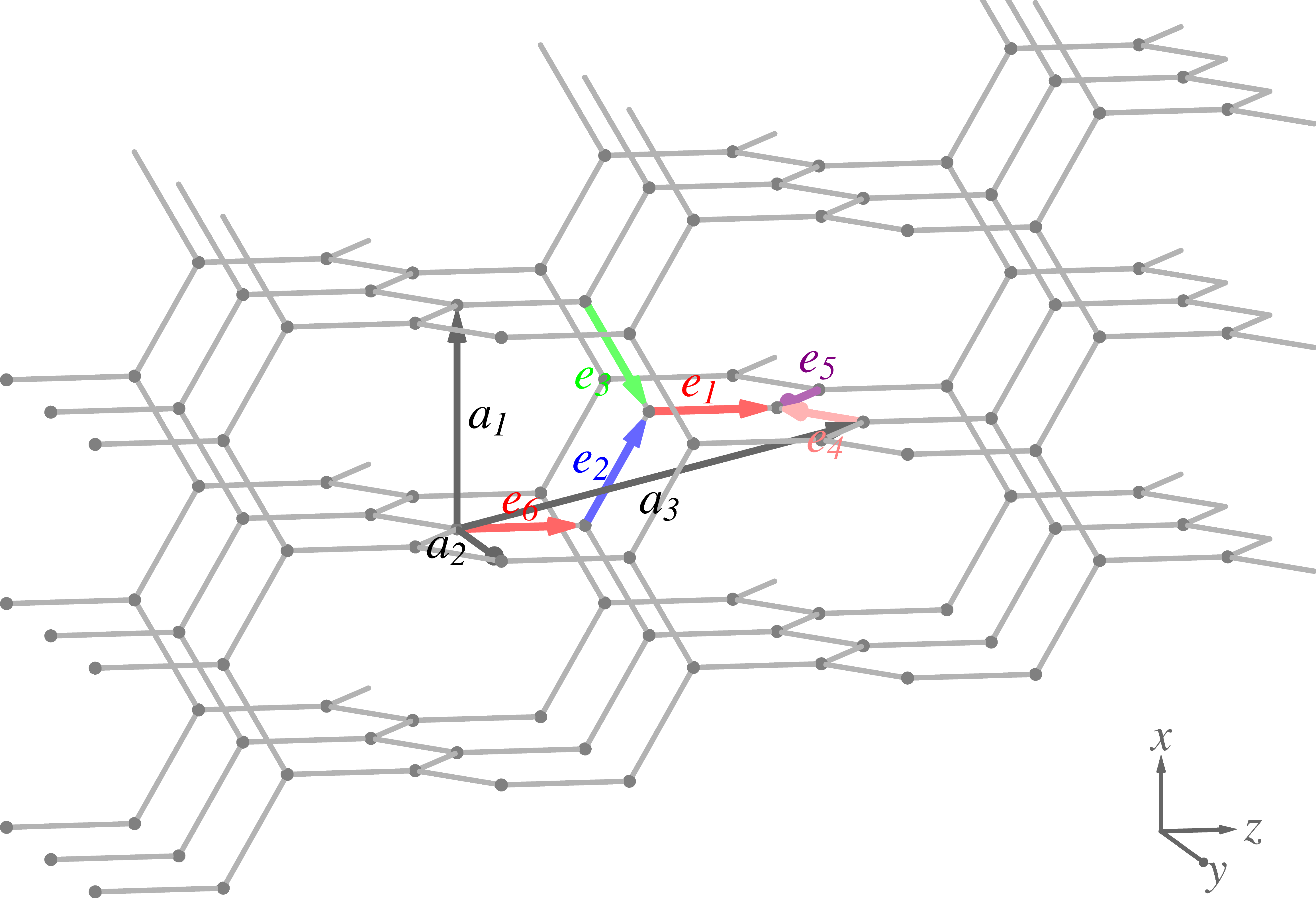}
  \caption{3D hyperhoneycomb lattice. The black arrows label the three primitive vectors $a_i$ while the six colored links $e_i$ form a unit cell.}
  \label{fig:hhc}
\end{figure}

To define the Hamiltonian of a gauge theory, one needs to make a gauge choice. The Kogut-Susskind Hamiltonian is constructed in temporal gauge. In this gauge, the Hamiltonian density for an SU(2) pure gauge theory in the continuum is given by
\begin{align}
    \mathcal{H}(\boldsymbol{x}) = \frac{1}{2} E_i^a(\boldsymbol{x})E_i^a(\boldsymbol{x}) + \frac{1}{2}B_i^a(\boldsymbol{x})B_i^a(\boldsymbol{x}) \,,
\end{align}
where the first term is the electric energy density and the second the magnetic energy density. The argument $\boldsymbol{x}$ labels a spatial point, the subscript $i$ is a spatial index and the superscript $a$ denotes an adjoint representation index. Repeated indices are implicitly summed over. As a consequence of choosing the temporal gauge, Gauss's law must be imposed at each vertex
\begin{align}
    (D_i E_i)^a(\boldsymbol{x}) = 0 \,, \quad \forall \boldsymbol{x}\,,a \,,
\end{align}
where $D_i$ denotes the covariant derivative. Physically, this means that the links joining at each vertex form an SU(2) singlet. When this Hamiltonian is discretized on a spatial lattice, trivalent lattices have the advantage that spatial gauge singlet states can be uniquely represented by the irreducible representation labels $j$ for each link without the need for any additional labels. This advantage has been exploited in previous studies for the quasi-1D plaquette chain~\cite{Klco:2019evd,Hayata:2021kcp,ARahman:2021ktn,ARahman:2022tkr}, the 2D honeycomb lattice~\cite{Muller:2023nnk} (see Ref.~\cite{Hayata:2023puo} for a $q$-deformed version), and the 3D triamond lattice~\cite{Kavaki:2024ijd,Kavaki:2025hcu}. In this work, we consider the hyperhoneycomb lattice as shown in Fig.~\ref{fig:hhc}, on which the KS Hamiltonian has been constructed~\cite{Illa:2025dou}. A unit cell of the hyperhoneycomb lattice is labeled by the six colored links in the figure. They specify five different directions
\begin{align}
    e_{1,6} &= (0,0,1)\,, ~~ e_2 = \frac{1}{2}(\sqrt{3},0,1)\,, ~~ e_3 = \frac{1}{2}(-\sqrt{3},0,1)\,, \nn
    \\
    e_4 &= \frac{1}{2}(0,-\sqrt{3},-1)\,, ~~ e_5 = \frac{1}{2}(0,\sqrt{3},-1) \,.
\end{align}
The whole lattice is generated by translationally repeating the unit cell along three directions that are specified by three primitive vectors
\begin{align}
    a_1 &= (\sqrt{3},0,0) \,, ~~ a_2 = (0,\sqrt{3},0)\,, ~~ a_3 = \frac{1}{2}(\sqrt{3},\sqrt{3},6) \,,
\end{align}
which are indicated by the three black arrows in the figure. The lattice spacing is specified by the edge length, which we take to be $1$ throughout. The volume of a unit cell is $V_{\rm cell} = |(a_1\times a_2)\cdot a_3| = 9$.

\begin{figure}[h]
  \centering
  \includegraphics[width=0.95\linewidth]{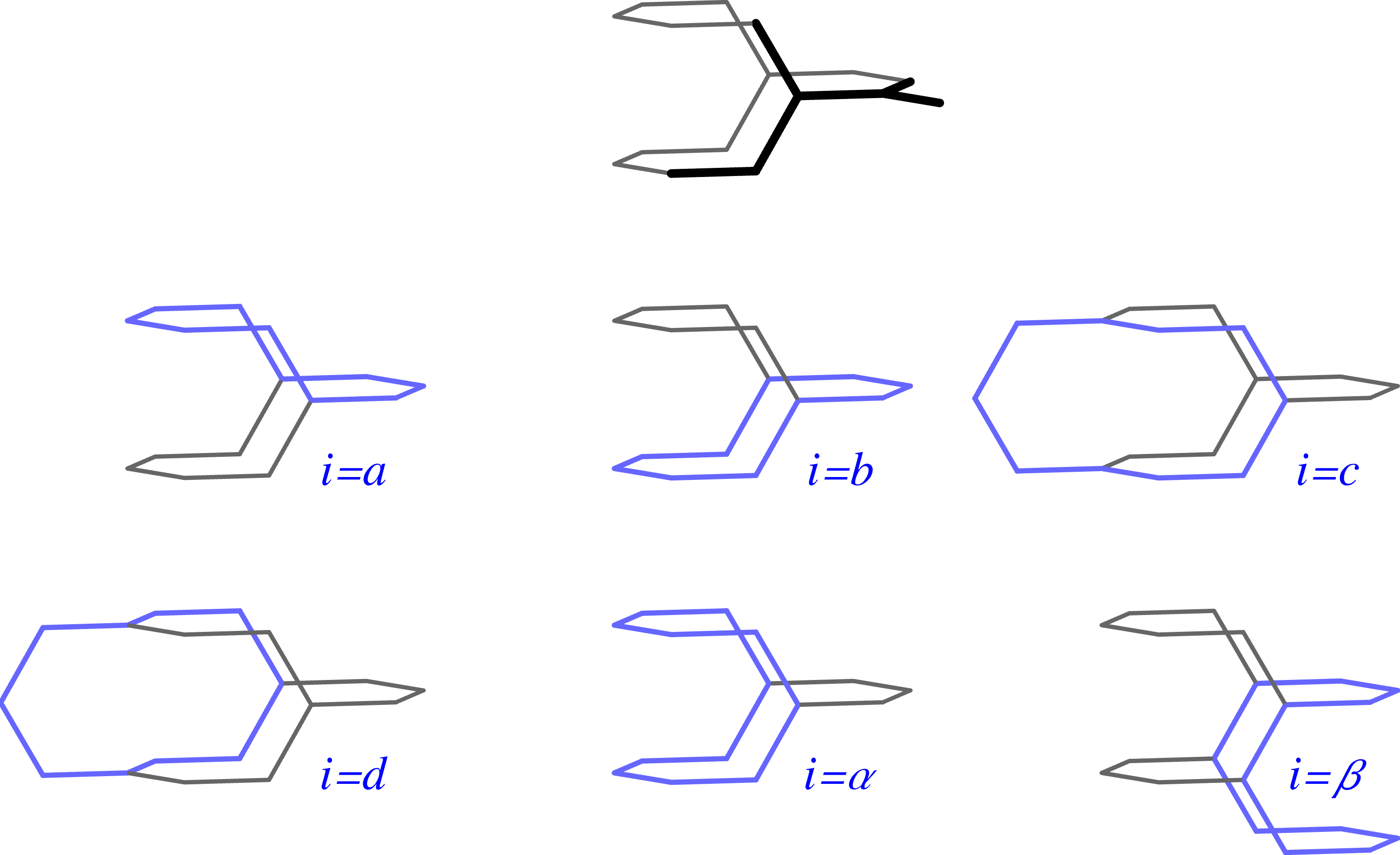}
  \caption{Shapes of six different Wilson loops (blue) used as plaquette operators in the magnetic Hamiltonian of a given unit cell, with the corresponding unit cell marked in black in the top row. }
  \label{fig:loops}
\end{figure}

The KS Hamiltonian on the hyperhoneycomb lattice can be written as~\cite{Illa:2025dou}
\begin{align}
    H &= H_E + H_B \,, \nn\\
    H_E &= 9 g^2\sum_{\rm cells}\bigg[\frac{2}{3}\sum_{\alpha\ne1,6} \mathrm{Tr}[(E_{\alpha})^2] +\frac{1}{6}\sum_{\alpha=1,6} \mathrm{Tr}[(E_{\alpha})^2] \bigg] \,, \nn\\
    H_B &= -\frac{4}{3g^2}\sum_{\rm cells}\bigg(\frac{1}{4} \sum_{i=a,b,c,d}P_{10}^i+\frac{5}{8}\sum_{i=\alpha,\beta} P_{12}^{i} \bigg) \,,
\end{align}
where we have neglected an overall constant in the magnetic term that depends only on the size of the lattice. Here $E_\alpha$ in the electric term denotes the electric field operator on a link $\alpha$ of a unit cell, and $P_{10}^i$ and $P_{12}^i$ in the magnetic term stand for plaquette operators made up of $10$-link and $12$-link Wilson loops, respectively. The shapes of these Wilson loops with different superscript labels are shown in Fig.~\ref{fig:loops}.

On this lattice, the Gauss's law constraint at each vertex $V$ is given by
\begin{align}
    \sum_{\alpha \in \ell(V)}E_\alpha = 0\,,
\end{align}
where $\ell(V)$ denotes all edges (links) connected to the vertex $V$.

Next we specify the matrix elements of the lattice Hamiltonian in the electric basis, labeled by the irreducible representation $j$. This is new compared with the previous work of Ref.~\cite{Illa:2025dou}. On trivalent lattices, one can completely solve Gauss's law at each vertex in the sense of integrating out the ``$m$'' quantum numbers for each $j$ and expressing the physical Hilbert space in terms of $j$ labels only. For a link $\alpha$, the electric energy term is diagonal in this basis
\begin{align}
\label{eqn:electric_link}
    \bra{J}\mathrm{Tr}[(E_{\alpha})^2]\ket{j} = \frac{1}{2}j(j+1)\delta_{jJ} \,.
\end{align}
The matrix elements of various plaquette operators can be obtained by generalizing the results in Refs.~\cite{Klco:2019evd,Muller:2023nnk}. We find 
\begin{align}
\label{eqn:plaq}
    & \bra{\{J\}} P_{10} \ket{\{j\}} = \prod_{V=1}^{10} \bigg[(-1)^{j_V+J_{V\oplus_{10}1}+j_{V,{\rm ext}}} \nn\\
    & \sqrt{(2J_V+1)(2j_{V\oplus_{10}1}+1)} \left\{\begin{matrix} j_{V,{\rm ext}} & j_V & j_{V\oplus_{10}1}
    \\
    \frac{1}{2} & J_{V\oplus_{10}1} & J_V\end{matrix} \right\} \bigg] \,, \nn\\
    & \bra{\{J\}} P_{12} \ket{\{j\}} = \prod_{V=1}^{12} \bigg[(-1)^{j_V+J_{V\oplus_{12}1}+j_{V,{\rm ext}}} \nn\\
    & \sqrt{(2J_V+1)(2j_{V\oplus_{12}1}+1)} \left\{\begin{matrix} j_{V,{\rm ext}} & j_V & j_{V\oplus_{12}1}
    \\
    \frac{1}{2} & J_{V\oplus_{12}1} & J_V\end{matrix} \right\} \bigg] \,,
\end{align}
where $V$ runs over all vertices along the active Wilson loop and we have defined $a\oplus_N b \equiv (a+b)\! \mod N$. Each vertex $V$ has three links connected to it. The subscripts $V$ and $V\oplus_N1$ denote the two links of the active plaquette under consideration, i.e., internal links, while the subscript $V,{\rm ext}$ denotes the external link that does not belong to the active plaquette. The states before and after the application of the plaquette operator are labeled by $\{j\}$ and $\{J\}$, respectively. For example, for the $P_{10}$ operator, $\{j\}$ stands for $\{j_1,j_2,\dots,j_{10}\}$. The curly brackets containing six numbers that are listed in two rows in the above equations are Wigner's $6j$ symbols.

\section{Minimal Truncation and Map onto Qubits}
\label{sec:H_onehalf}
\subsection{Electric Flux Loop Basis and Qubit Representation}
In the remainder of this study, we focus on the minimal truncation in the electric basis $j_{\rm max}=1/2$, meaning that each link can be in either the $j=0$ or $j=1/2$ state. On trivalent lattices, Gauss's law can be completely solved: The only allowed states at each trivalent vertex are $\{\ket{0,0,0},\ket{0,1/2,1/2},\ket{1/2,0,1/2},\ket{1/2,1/2,0}\}$. This results in a physical Hilbert space made up of electric flux loops of various shapes under the minimal truncation. These electric flux loops can be generated by applying a set of plaquette operators to the strong coupling vacuum state, in which all links are in the $j=0$ state. We want to find the minimal set of these plaquette operators that can be used to specify the basis of the physical Hilbert space.

A naive starting point to find this set is the union of all the six plaquette operators as shown in Fig.~\ref{fig:loops} across the whole lattice. These six plaquette operators are associated with the unit cell shown in black in the figure. If the unit cell is at position $(x,y,z)$ along the $a_1,a_2,a_3$ directions, i.e., at $xa_1+ya_2+za_3$, $x,y,z\in\mathbb{Z}$, then we use the same position to label all the six plaquette operators. The state at $(x,y,z)$ can be specified by six binary numbers, each of which represents whether the corresponding plaquette operator has been applied or not. For example, $|0\rangle$ means the corresponding plaquette operator has not been applied, while $|1\rangle$ means it has been applied. If we map these binary number states onto qubits, the relevant plaquette operators simply correspond to Pauli-$X$ matrices, up to some prefactors that do not flip the qubits. We use the notation $\sigma^x_{xyzi}$ for these plaquette operators (up to the prefactors that we will specify later), where the subscript $xyz$ specifies the position of the unit cell with which the six plaquette operators are associated and $i=a,b,c,d,\alpha,\beta$ labels each of the six.

Under the minimal truncation, not all these plaquette operators generate independent states. For the six plaquette operators associated with one unit cell, we have the following constraints
\begin{align}
\label{eqn:constraint1}
    \sigma^{x}_{xyz\alpha}&=\sigma^{x}_{xyza}\sigma^{x}_{xyzb} \,, \nn\\
    \sigma^{x}_{xyz\beta}&=\sigma^{x}_{xyzb}\sigma^{x}_{(x-1)yza} \,, \nn\\
    \sigma^{x}_{xyzd}&=\sigma^{x}_{xyza}\sigma^{x}_{xyzb}\sigma^{x}_{xyzc} \,.
\end{align}
Furthermore, across the lattice we have another constraint
\begin{align}
\label{eqn:constraint2}
    \sigma^{x}_{x(y-1)zc}\sigma^{x}_{xyzd} = \sigma^{x}_{(x+1)y(z-1)\beta} \,.
\end{align}
Under these constraints, we find that the set of independent basis states can be generated by the operators
\begin{align}
    \sigma^{x}_{xyza}\,,~~ \sigma^{x}_{xyzb}\,,~~\sigma^{x}_{x0zc} \,.
\end{align}
The type-$d$, -$\alpha$, and -$\beta$ plaquette operators are related to these via Eq.~\eqref{eqn:constraint1}. The type-$c$ plaquette operators at $y\neq0$ are obtained iteratively via
\begin{align}
&\sigma^x_{x(y+1)zc} \nn\\
\label{eqn:c-type}
&=\sigma^x_{xyzc} \sigma^x_{x(y+1)za}\sigma^x_{x(y+1)zb}\sigma^x_{(x+1)(y+1)(z-1)b}\sigma^x_{x(y+1)(z-1)a} \,,
\end{align}
where we have used Eqs.~\eqref{eqn:constraint1} and~\eqref{eqn:constraint2}.

\begin{figure}[t]
    \centering
    \includegraphics[width=0.9\linewidth]{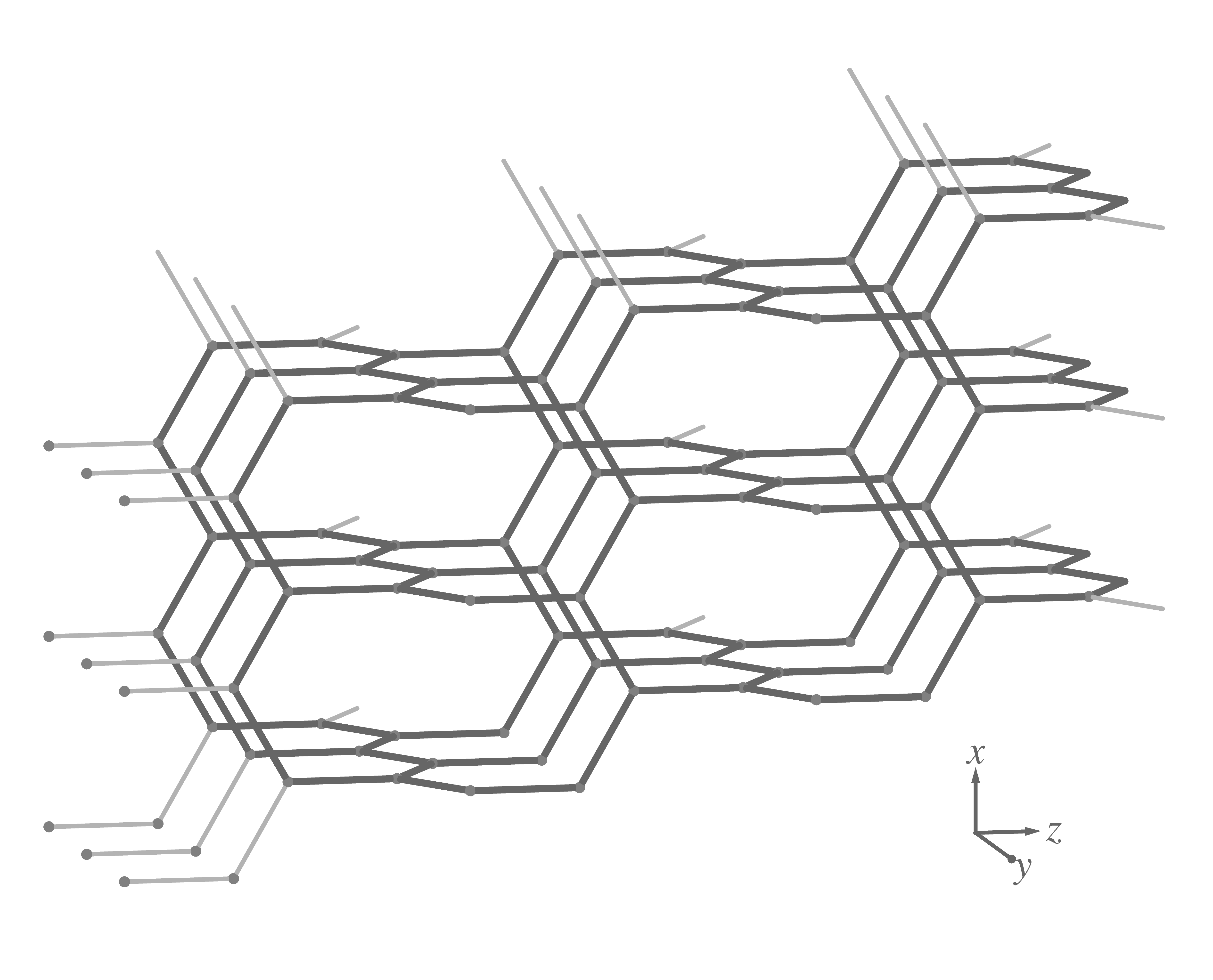}
    \caption{$3\times3\times3$ hyperhoneycomb lattice under fixed boundary conditions where all outside links are in the $j=0$ state.
    The light gray links represent all links forced to be in the $j=0$ state due to the boundary conditions and Gauss's law. Only the dark gray links can be excited, i.e., be in the $j=0$ and $j=1/2$ states.}
    \label{fig:3b3b3}
\end{figure}

For a $n_x\times n_y\times n_z$ hyperhoneycomb lattice, by which we mean that a unit cell is repeated $n_i$ times along the $a_i$ direction, we have $x\in \mathbb{Z}_{n_x}$, $y\in \mathbb{Z}_{n_y}$ and $z\in \mathbb{Z}_{n_z}$. The total number of qubits needed to specify the physical Hilbert space is roughly
\begin{align}
    N_q \approx 2n_xn_yn_z + n_xn_z \,,
\end{align}
where the prefactor $2$ originates from summing over $a$ and $b$ types of loops and $n_xn_z$ comes from the $c$ type at $y=0$.
This is a rough estimate. The exact number depends on the boundary conditions of the hyperhoneycomb lattice. Under periodic boundary conditions, the Hilbert space of this dual lattice consists of more than one copy of the physical Hilbert space of the original truncated SU(2) LGT. These extra copies are coupled by a collection of qubit flips. For example, the product of the type-$c$ qubit flips on the $y=0$ plane, $\prod_{x,z}\sigma_{x0zc}^x$, is proportional to an identity operator in the original truncated SU(2) LGT. Under open boundary conditions, or more specifically, fixed boundary conditions where all links outside the lattice are in the $j=0$ state, the qubit cost is illustrated in Table~\ref{tab:qubit-convention}. Due to the boundary conditions, some links of the unit cells on the boundary are forced to be in the $j=0$ state and the plaquette operators acting on them do not exist. Furthermore, the type-$\beta$ loops at $z=0$ become independent since there are no $a$ and $b$ types to generate them according to Eq.~\eqref{eqn:constraint1}. An example of a $3\times3\times3$ hyperhoneycomb lattice is depicted in Fig.~\ref{fig:3b3b3}. Collecting the results in Table~\ref{tab:qubit-convention}, we see that the total number of qubits needed is
\begin{widetext}
\begin{align}
\label{eqn:qubit_count}
    N_q = \underbrace{2\, (n_x-1) (n_y-1) (n_z-1)}_{\text{bulk}}
  \,+\, \underbrace{(n_x-1) (n_y-1) + (n_y-1) (n_z-1) + (n_z-1) (n_x-1)}_{\text{three boundary planes}} \,.
\end{align}
To decompose the total qubit count into the bulk and boundary contributions, we have written $n_x$ as $(n_x-1)+1$ for the type-$b$ loops in Table~\ref{tab:qubit-convention}.

\begin{table}[h]
  \centering
  \begin{tabular}{c|l|c}
    \toprule
    Type & Cells $(x,y,z)$ carrying independent qubits & Multiplicity \\
    \midrule
    $a$  & $0 \le x < n_x-1$,\ $1 \le y < n_y$,\ $1 \le z < n_z$,   & $(n_x-1) (n_y-1) (n_z-1)$ \\
    $b$  & $0 \le x < n_x$,\ $1 \le y < n_y$,\ $1 \le z < n_z$,   & $n_x(n_y-1)(n_z-1)$ \\
    $c$  & $0 \le x < n_x-1$,\ $y = 0$,\ $1 \le z < n_z$,  & $(n_x-1)(n_z-1)$ \\
    $\beta$  & $1 \le x < n_x$ ,\ $1 \le y < n_y$,\ $z = 0$,  & $(n_x-1)(n_y-1)$ \\
    \bottomrule
  \end{tabular}
  \caption{Qubit assignment for the spatial-gauge-singlet basis states on the hyperhoneycomb lattice with fixed boundary conditions where all outside links are in the $j=0$ state. All four cases are independent of each other.}
  \label{tab:qubit-convention}
\end{table}

\subsection{Hamiltonian in Qubit Representation}
We are now ready to explicitly write out the lattice Hamiltonian in terms of Pauli matrices acting on the independent qubits as specified in the previous subsection.

The electric energy of a unit cell consists of contributions from six links, as shown in Fig.~\ref{fig:hhc}. If a link is excited to the $j=1/2$ state, it contributes $3/8$ according to Eq.~\eqref{eqn:electric_link}. When in the $j=0$ state, the link does not contribute to the electric energy. With these, we can write the total electric energy term in the Hamiltonian as
\begin{align}
    H_E &= \frac{27g^2}{8}\sum_{xyz}\bigg[\frac{2}{3}(\mathcal{O}'^{2}_{xyz}+\mathcal{O}'^{3}_{xyz}+\mathcal{O}'^{4}_{xyz}+\mathcal{O}'^{5}_{xyz}) + \frac{1}{6}(\mathcal{O}'^{1}_{xyz}+\mathcal{O}'^{6}_{xyz}) \bigg]\,,
    \label{eqn:HE_spin}
\end{align}
where $\mathcal{O}'^{\alpha}_{xyz}$s are operators that return $1$ if the link $\alpha$ at position $(x,y,z)$ is excited to the $j=1/2$ state and $0$ if it is in the $j=0$ state. More specifically, they can be written as
\begin{align}
    \mathcal{O}'^{\alpha}_{xyz}=\frac{1-\mathcal{O}^{\alpha}_{xyz}}{2}\,,
\end{align}
with 
\begin{align}
\mathcal{O}_{xyz}^{1}&=\sigma^z_{xyzb}\,\sigma^z_{xyza}\,\sigma^z_{x(y+1)zb}\,\sigma^z_{x(y+1)za}\,\sigma^z_{xy(z+1)c}\,\sigma^z_{(x-1)y(z+1)c} \,,\nn \\    \mathcal{O}_{xyz}^{2}&=\sigma^z_{xyzb}\,\sigma^z_{x(y+1)zb}\,\sigma^z_{xyzc}\,\sigma^z_{(x-1)y(z+1)c}\,, \nn \\
\mathcal{O}_{xyz}^{3}&=\sigma^z_{xyza}\,\sigma^z_{x(y+1)za}\,\sigma^z_{xyzc}\,\sigma^z_{xy(z+1)c} \,, \nn\\
\mathcal{O}_{xyz}^{4}&=\sigma^z_{x(y+1)zb}\,\sigma^z_{x(y+1)za}\,\sigma^z_{xy(z+1)c}\,\sigma^z_{(x-1)y(z+1)c}\,\sigma^z_{xy(z+1)b}\,\sigma^z_{(x-1)y(z+1)a}\,, \nn \\
\mathcal{O}_{xyz}^{5}&=\sigma^z_{xyzb}\,\sigma^z_{xyza}\,\sigma^z_{xy(z+1)b}\,\sigma^z_{(x-1)y(z+1)a} \,, \nn\\
\mathcal{O}_{xyz}^{6}&=\sigma^z_{xyzb}\,\sigma^z_{(x-1)yza}\,\sigma^z_{x(y+1)zb}\,\sigma^z_{(x-1)(y+1)za}\,\sigma^z_{xyzc}\,\sigma^z_{(x-1)yzc}\,, 
\end{align}
where $\sigma^z_{xyzc}$ satisfies a similar relation as $\sigma^x_{xyzc}$ in Eq.~\eqref{eqn:c-type}. The operator $\mathcal{O}_{xyz}^{\alpha}$ is the product of $\sigma^z$s over the independent plaquettes that flip the state on link $\alpha$ (four or six plaquettes for each link). Because an excited plaquette corresponds to $\sigma^z=-1$, $\mathcal{O}_{xyz}^{\alpha}$ measures the parity of the number of excited plaquettes on link $\alpha$: It equals $-1$ if the number is odd and $+1$ if even.

The magnetic energy part of the Hamiltonian can be written as
\begin{align}
    H_B&= - \frac{1}{3g^2b}\sum_{xyz}\bigg[\sum_{i=a,b,c} (-0.5)^{c_{xyzi}}\sigma_{xyzi}^x+(-0.5)^{c_{xyzd}}\sigma_{xyza}^x\sigma_{xyzb}^x\sigma_{xyzc}^x
    \nn \\
    &\qquad +\frac{5}{2}(-0.5)^{c_{xyz\alpha}}\sigma_{xyza}^x\sigma_{xyzb}^x + \frac{5}{2}(-0.5)^{c_{xyz\beta}}\sigma_{xyzb}^x\sigma_{(x-1)yza}^x \bigg] \,, \label{eqn:HB_spin}
\end{align}
where we have used Eq.~\eqref{eqn:constraint1} for the type-$d$, -$\alpha$, and -$\beta$ plaquette operators. The prefactors $(-0.5)^{c_{xyzi}}$ denote controls that depend on the qubit states. They originate from the SU(2) Clebsch-Gordan coefficients when we solve Gauss's law and appear in the generic matrix elements of plaquette operators as Wigner's $6j$ symbols. With the truncation $j_{\rm max}=1/2$, one can evaluate Eq.~\eqref{eqn:plaq} and obtain these prefactors: If exactly $k$ external links attached to the active plaquette under consideration are excited to the $j=1/2$ state, we have $c_{xyzi}=k/2$. Writing these $c_{xyzi}$ symbols out explicitly as operators leads to
\begin{align}
c_{xyza}&=\frac{1}{2}[\mathcal{O}'^{5}_{(x+1)(y+1)(z-1)}+\mathcal{O}'^{2}_{(x+1)yz}+\mathcal{O}'^{2}_{xyz}+\mathcal{O}'^{4}_{xyz}+\mathcal{O}'^{6}_{x(y-1)(z+1)} +\mathcal{O}'^{5}_{x(y-1)z}+\mathcal{O}'^{2}_{x(y-1)z}+\mathcal{O}'^{2}_{(x+1)(y-1)z} \nn
\\
&\qquad +\mathcal{O}'^{4}_{(x+1)(y-1)(z-1)}+\mathcal{O}'^{1}_{(x+1)y(z-1)}] \,,\nn
\\
c_{xyzb}&=\frac{1}{2}[\mathcal{O}'^{5}_{x(y+1)(z-1)}+\mathcal{O}'^{3}_{(x-1)yz}+\mathcal{O}'^{3}_{xyz}+\mathcal{O}'^{4}_{xyz}+\mathcal{O}'^{6}_{x(y-1)(z+1)} +\mathcal{O}'^{5}_{x(y-1)z}+\mathcal{O}'^{3}_{x(y-1)z}+\mathcal{O}'^{3}_{(x-1)(y-1)z} \nn
\\
&\qquad +\mathcal{O}'^{4}_{x(y-1)(z-1)}+\mathcal{O}'^{1}_{xy(z-1)}] \,,\nn
\\
c_{xyzc}&=\frac{1}{2}[\mathcal{O}'^{5}_{x(y+1)(z-1)}+\mathcal{O}'^{3}_{(x-1)yz}+\mathcal{O}'^{1}_{xyz}+\mathcal{O}'^{2}_{(x+1)yz}+\mathcal{O}'^{5}_{(x+1)(y+1)(z-1)} +\mathcal{O}'^{5}_{(x+1)y(z-1)}+\mathcal{O}'^{3}_{(x+1)y(z-1)} \nn
\\
&\qquad +\mathcal{O}'^{6}_{(x+1)y(z-1)} +\mathcal{O}'^{2}_{xy(z-1)}+\mathcal{O}'^{5}_{xy(z-1)}] \,,\nn
\\
c_{xyzd}&=\frac{1}{2}[\mathcal{O}'^{4}_{x(y-1)(z-1)}+\mathcal{O}'^{3}_{(x-1)(y-1)z}+\mathcal{O}'^{1}_{x(y-1)z}+\mathcal{O}'^{2}_{(x+1)(y-1)z}+\mathcal{O}'^{4}_{(x+1)(y-1)(z-1)} +\mathcal{O}'^{4}_{(x+1)y(z-1)} \nn
\\
&\qquad +\mathcal{O}'^{3}_{(x+1)y(z-1)} +\mathcal{O}'^{6}_{(x+1)y(z-1)}+\mathcal{O}'^{2}_{xy(z-1)}+\mathcal{O}'^{4}_{xy(z-1)}] \,,\nn
\\
c_{xyz\alpha}&=\frac{1}{2}[\mathcal{O}'^{5}_{x(y+1)(z-1)}+\mathcal{O}'^{3}_{(x-1)yz}+\mathcal{O}'^{1}_{xyz}+\mathcal{O}'^{2}_{(x+1)yz}+\mathcal{O}'^{5}_{(x+1)(y+1)(z-1)} + \mathcal{O}'^{1}_{(x+1)y(z-1)} +\mathcal{O}'^{4}_{(x+1)(y-1)(z-1)} \nn
\\
&\qquad +\mathcal{O}'^{2}_{(x+1)(y-1)z}+\mathcal{O}'^{1}_{x(y-1)z}+\mathcal{O}'^{3}_{(x-1)(y-1)z}
+\mathcal{O}'^{4}_{x(y-1)(z-1)}+\mathcal{O}'^{1}_{xy(z-1)}] \,, \nn
\\
c_{xyz\beta}&=\frac{1}{2}[\mathcal{O}'^{6}_{xyz}+\mathcal{O}'^{3}_{xyz}+\mathcal{O}'^{4}_{xyz}+\mathcal{O}'^{6}_{x(y-1)(z+1)}+\mathcal{O}'^{5}_{x(y-1)z} + \mathcal{O}'^{3}_{x(y-1)z}+\mathcal{O}'^{6}_{x(y-1)z}+\mathcal{O}'^{2}_{(x-1)(y-1)z} \nn
\\
&\qquad +\mathcal{O}'^{5}_{(x-1)(y-1)z}+\mathcal{O}'^{6}_{(x-1)(y-1)(z+1)} +\mathcal{O}'^{4}_{(x-1)yz}+\mathcal{O}'^{2}_{(x-1)yz}] \,.
\end{align}
To convert the control prefactors into products of Pauli matrices, we use
\begin{align}
    (-0.5)^{c_{xyzi}} &= \prod_{k}\sqrt{-0.5}^{\mathcal{O}'_k} = \prod_{k}\left[\left(\frac{1}{2}-\frac{i}{2\sqrt{2}}\right)\mathcal{O}_k+\left(\frac{1}{2}+\frac{i}{2\sqrt{2}}\right)\right] \,,
\end{align}
where we have chosen $\sqrt{-0.5}=i/\sqrt{2}$ and $k$ runs over all $\mathcal{O}'$ operators appearing in $c_{xyzi}$.

Putting everything together, the Hamiltonian in the physical sector under the minimal truncation can be written as a sum of Pauli strings of various lengths. One can further decompose the Hamiltonian evolution operator into CNOT gates and single-qubit rotations using Trotterization. We leave this to future work, since for fault-tolerant quantum computing the compilation will be completely different.
\end{widetext}

\section{Quantum Ergodicity and Eigenstate Thermalization}
\label{sec:results}
In this section we will investigate whether the hyperhoneycomb lattice system under the minimal truncation obeys the eigenstate thermalization hypothesis (ETH).
The ETH states that for a non-integrable Hamiltonian, the diagonal matrix elements of a local observable $O$ in the energy eigenbasis are very close to its microcanonical expectation value $O_{\rm mc}$, while the off-diagonal ones are exponentially suppressed
\begin{equation}
    \langle E_m|O|E_n\rangle = O_{\rm mc}(\bar E)\,\delta_{mn} + e^{-S(\bar E)/2} f_O(\bar E,\omega)\,R_{mn} \,,
    \label{eqn:eth}
\end{equation}
where the two eigenstates are labeled by $|E_m\rangle$ and $|E_n\rangle$ with energies $E_m$ and $E_n$, respectively, $\bar E=(E_m+E_n)/2$, $\omega=E_m-E_n$, $S(\bar E)$ is the thermodynamic entropy, $f_O$ is a well-behaved function of $\bar{E}$ and $\omega$, and $R_{mn}$ is a close-to-Gaussian~\cite{Murthy:2019fgs} random variable with zero mean and unit variance. For a detailed review, see e.g., Ref.~\cite{DAlessio2015}.

In order to test the ETH in the lattice theory, we will compute the matrix elements of some local operators on a hyperhoneycomb lattice of size $(n_x,n_y,n_z)=(2,3,3)$. Since the lattice is of finite size, we first need to identify the range of couplings in which the system is quantum ergodic, i.e., in which the lattice Hamiltonian is non-integrable.

\subsection{Level Statistics}
Our SU(2) Hamiltonian in the electric basis becomes integrable in the strong coupling limit $g\to \infty$ as only the electric energy term remains. For any finite $g\neq0$, exact integrability is broken and we expect the lattice system to be non-integrable when its volume is sufficiently large. To identify the coupling range of non-integrability or quantum ergodicity on a finite lattice, we analyze the statistics of energy levels.
For a time-reversal symmetric Hamiltonian, the level spacings are expected to follow the Wigner surmise with Dyson index $\beta=1$
\begin{equation}
\label{eqn:WD}
    P_{\beta=1}(s)=\frac{\pi}{2}s\, e^{-\frac{\pi}{4} s^2} \,,
\end{equation}
where $s$ is the nearest-neighbor spacing in the unfolded spectrum. Unfolding means rescaling the spectrum so that it has unit mean level spacing throughout. On the other hand, for an integrable system the level spacing statistics follow the Poisson distribution, $P_{\rm P}(s)=e^{-s}$.

\begin{figure}[ht]
    \centering
    \includegraphics[width=1\linewidth]{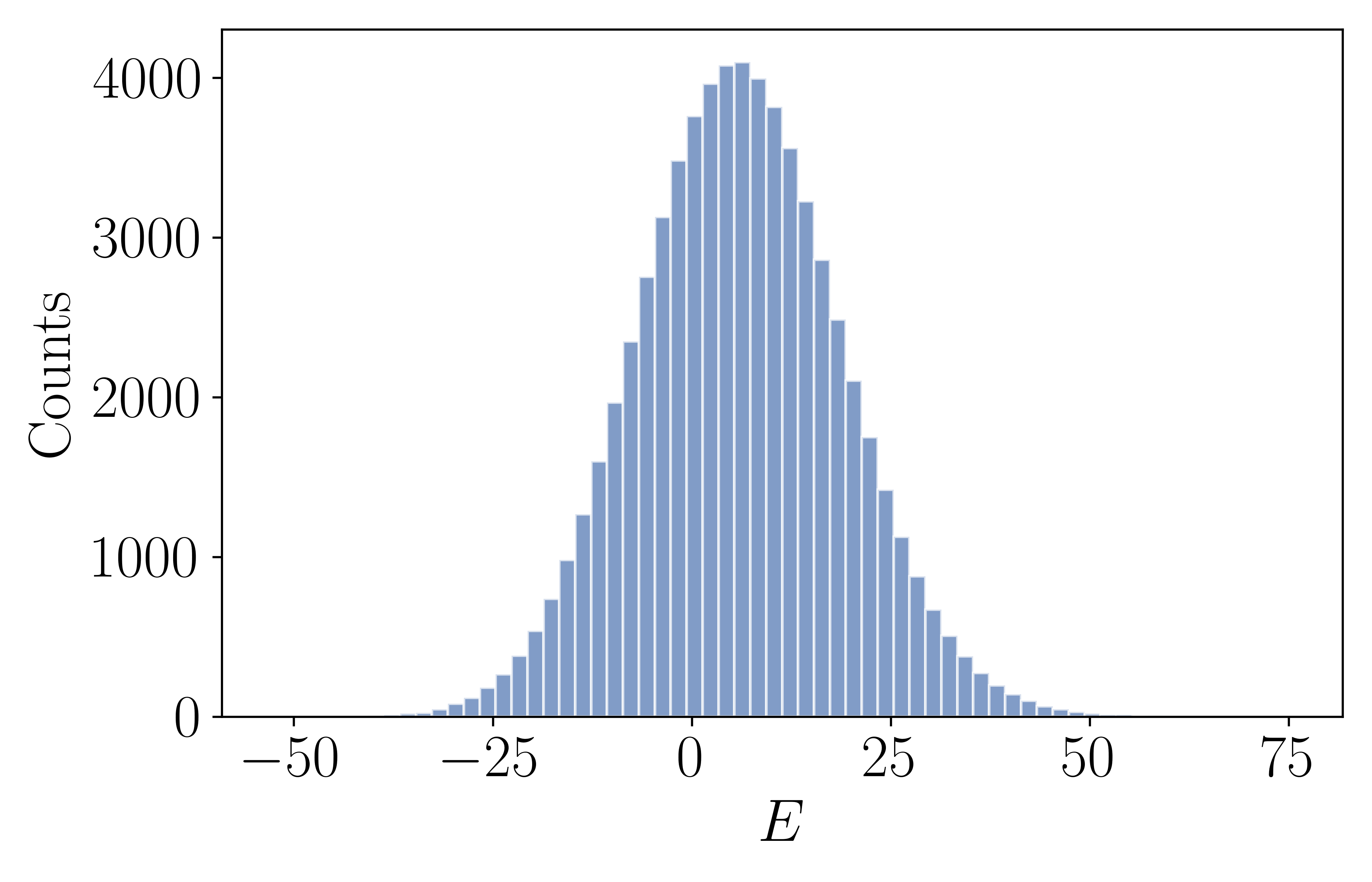}
    \caption{Density of states of the hyperhoneycomb lattice Hamiltonian with $(n_x,n_y,n_z)=(2,3,3)$ under fixed boundary conditions at $g=0.3$. In total, $2^{16}=65536$ energy levels are histogrammed with a bin size of $\Delta E=2$.
    }
    \label{fig:levels}
\end{figure}

We perform exact diagonalization for the Hamiltonian on a hyperhoneycomb lattice of size $(n_x,n_y,n_z)=(2,3,3)$ under fixed boundary conditions, which corresponds to $N_q=16$ qubits and $2^{16}=65536$ energy levels. 
At the lattice coupling $g=0.3$, a histogram of energy levels is shown in Fig.~\ref{fig:levels} with a bin size of $\Delta E=2$. Compared with the results of the previous study under the same SU(2) truncation on plaquette chains and honeycomb lattices~\cite{Ebner:2023ixq}, the density of states here is more symmetric with respect to the peak here.

\begin{figure}[ht]
    \centering
    \includegraphics[width=1\linewidth]{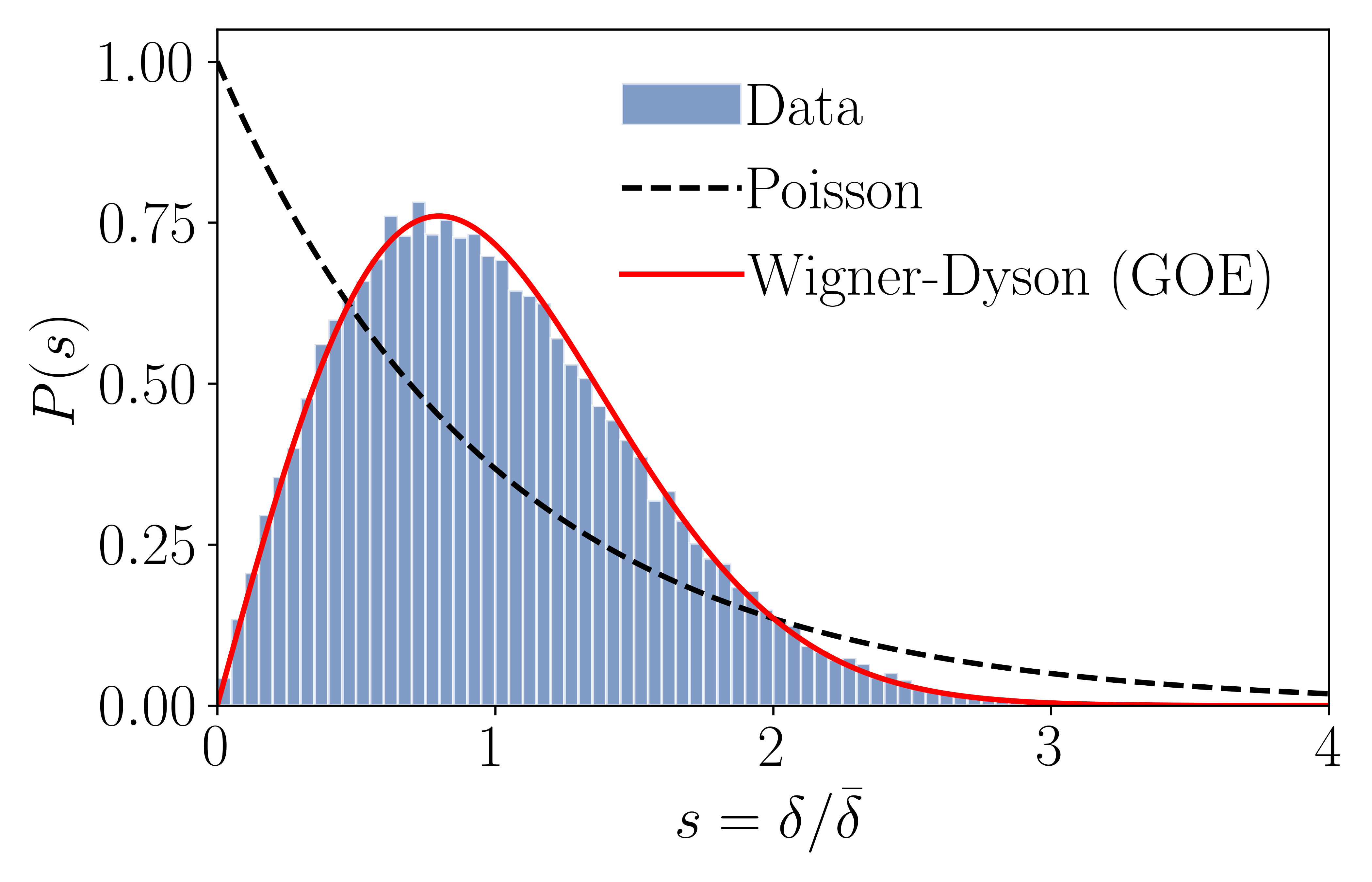}
    \caption{Level spacing statistics of the $(n_x,n_y,n_z)=(2,3,3)$ hyperhoneycomb lattice Hamiltonian under fixed boundary conditions at $g=0.3$. At each edge of the spectrum $10\%$ of the levels are discarded to reduce finite-size effects and the rest of the eigenvalues are unfolded to have unit mean. The dashed black line shows the Poisson distribution $P_{\rm P}(s)=e^{-s}$, while the solid red line shows the Wigner-Dyson distribution $P_{\beta=1}(s)=\frac{\pi}{2}s\, e^{-\frac{\pi}{4} s^2}$. The bin width of the histogram is $\Delta s=0.05$ here.
    }
    \label{fig:stat}
\end{figure}

In order to reduce finite-size effects in the level spacing analysis, we discard {10\%} of the eigenvalues at each edge of the spectrum and unfold the remaining eigenvalues such that $\langle s \rangle=1$. Specifically, we divide each level spacing $\delta_n=E_{n+1}-E_n$ by an estimated local mean spacing,
\begin{equation}
    s_n=\frac{\delta_n}{\langle\delta_n\rangle}, \qquad \langle\delta_n\rangle=\frac{1}{2k+1}\sum_{m=n-k}^{n+k}\delta_m \,,
\end{equation}
with $k=10$, and then normalize the spectrum by an overall constant to enforce $\langle s \rangle=1$. Figure~\ref{fig:stat} shows that the level statistics closely follow the Wigner-Dyson distribution~\eqref{eqn:WD} for the Gaussian Orthogonal Ensemble (GOE), featuring level repulsion near $s=0$. For comparison, the Poisson distribution is also shown. This close agreement implies that the lattice Hamiltonian with $(n_x,n_y,n_z)=(2,3,3)$ and fixed boundary conditions has no degeneracy in the spectrum and thus has no residual discrete symmetry that would need to be resolved.

We also study the restricted ratio of consecutive spacings
\begin{equation}
    r_n=\frac{\min(\delta_n,\delta_{n+1})}{\max(\delta_n,\delta_{n+1})}\,,
\end{equation}
as an indicator of quantum ergodicity. 
Its expected values are $\langle r\rangle_{\rm P}\approx 0.386$ for Poisson statistics and $\langle r\rangle_{\rm GOE}\approx 0.531$ for the GOE. At $g=0.3$, the calculated mean of the restricted gap ratio is $\langle r \rangle=0.529$, reconfirming that the $(n_x,n_y,n_z)=(2,3,3)$ hyperhoneycomb lattice Hamiltonian is quantum ergodic at this coupling.

\begin{figure}[ht]
    \centering
    \includegraphics[width=1\linewidth]{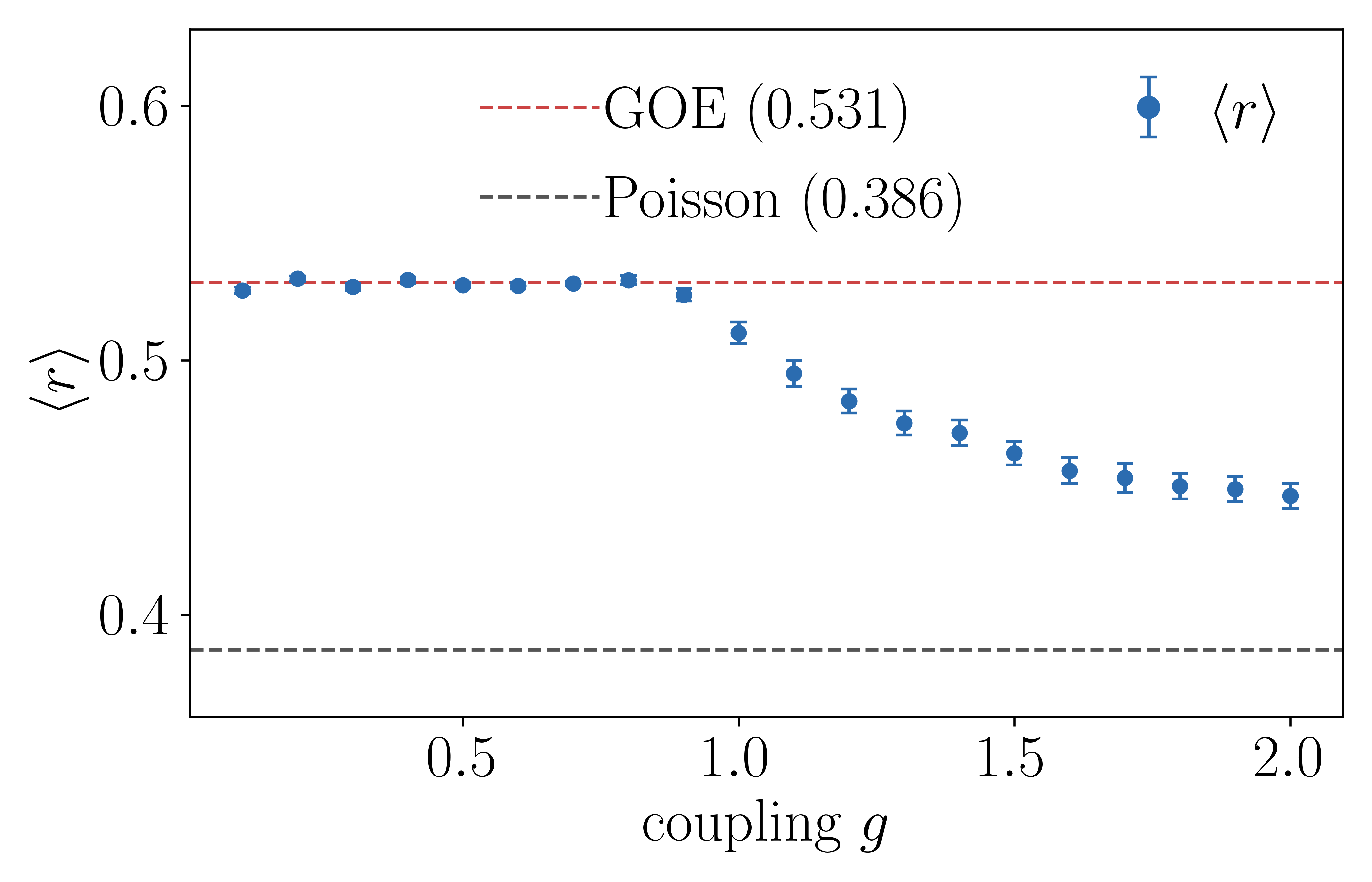}
    \caption{Mean restricted gap ratio $\langle r
    \rangle$ as a function of coupling $g$ for the $(n_x,n_y,n_z)=(2,3,3)$ hyperhoneycomb lattice Hamiltonian under fixed boundary conditions. The red and black dashed lines show the GOE and Poisson values, respectively. The error bars denote the standard uncertainty of the mean.
    }
    \label{fig:rvsg}
\end{figure}

We calculate the average restricted gap ratios for a range of different couplings in the same lattice Hamiltonian. The results are shown in Fig.~\ref{fig:rvsg}. It turns out that
the $(n_x,n_y,n_z)=(2,3,3)$ lattice Hamiltonian is quantum ergodic in the region $g\le0.8$. At larger couplings, $\langle r \rangle$ decreases towards the Poisson value $\langle r \rangle_{\rm P} = 0.386$, as the electric term starts to dominate the Hamiltonian. Because the electric term is diagonal in the electric basis we choose, the Hamiltonian at infinite coupling has symmetries given by $[H_E,\sigma_n^z]=0$ for each qubit $n$ describing the physical states. The number of conserved quantities equals that of the degrees of freedom. Thus, the lattice system becomes integrable in the infinite coupling limit. On the other hand, at any finite coupling, we expect the lattice theory to be quantum ergodic when the lattice size is sufficiently large. An explicit demonstration of this volume dependence is beyond the capability of exact diagonalization, since an $(n_x,n_y,n_z)=(3,3,3)$ lattice corresponds to $28$ qubits according to Eq.~\eqref{eqn:qubit_count}. However, this has been shown numerically, e.g., for SU(3) plaquette chains~\cite{Chen:2026hnh}.

\subsection{Matrix Elements of Local Observables}
To test the ETH directly, we choose the energy density operators across the lattice as the local observables and calculate their matrix elements in the eigenbasis of the $(n_x,n_y,n_z)=(2,3,3)$ hyperhoneycomb lattice system at the ergodic coupling $g=0.3$. The energy density operator at lattice site $(x,y,z)$ consists of electric and magnetic parts that can be read off from Eqs.~\eqref{eqn:HE_spin} and~\eqref{eqn:HB_spin}, respectively. 

The ETH ansatz~\eqref{eqn:eth} predicts that the diagonal matrix elements of a local operator approach its microcanonical ensemble average in the thermodynamic limit and depend only on the energy. The eigenstate-to-eigenstate fluctuations in the diagonal matrix elements are random and decrease exponentially with the system size, since the thermodynamic entropy $S(\bar{E})$ is extensive. Physically, this means that the highly excited eigenstates appear thermal when probed by a local observable. 

\begin{figure}[ht]
    \centering
    \includegraphics[width=1\linewidth]{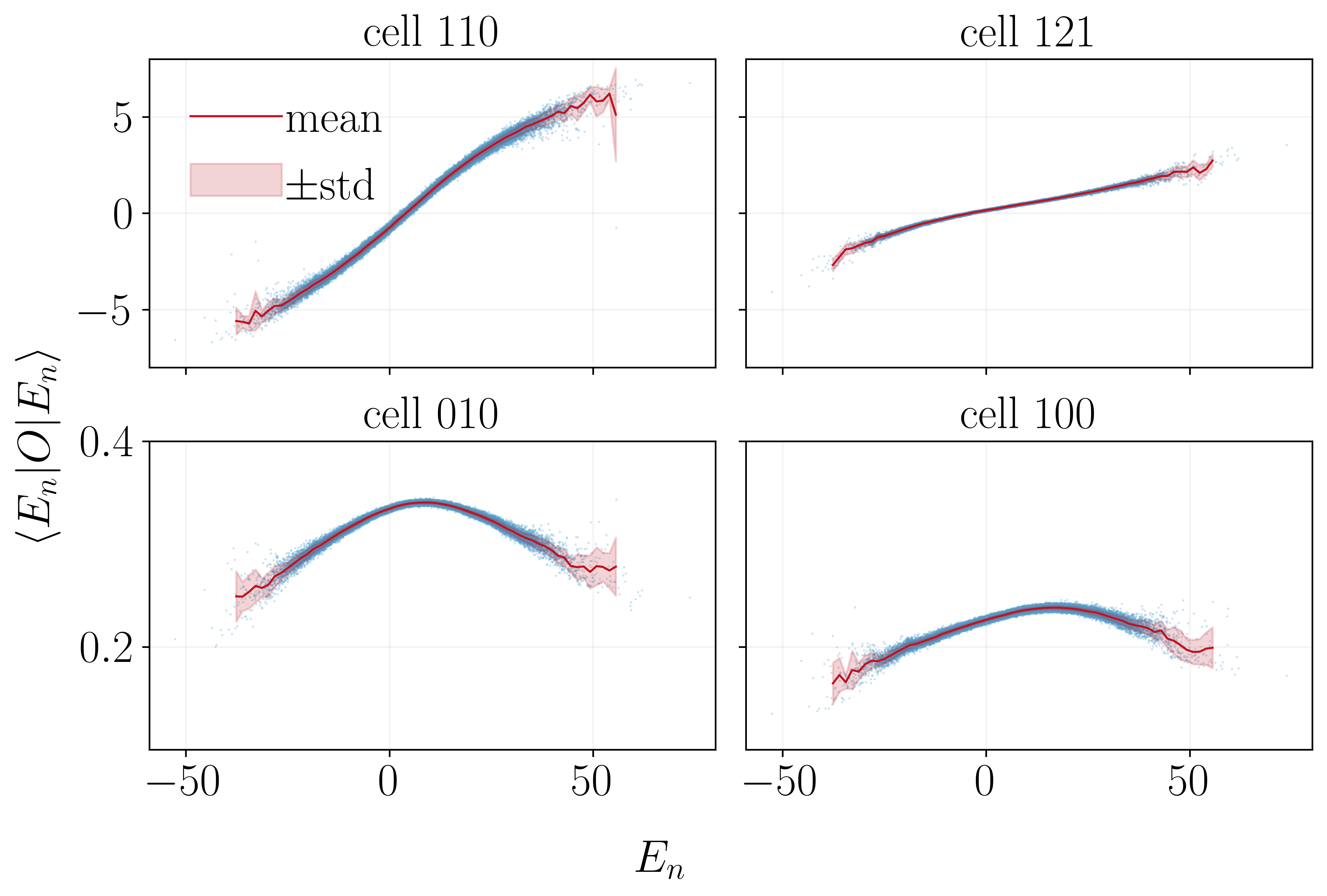}
    \caption{Diagonal matrix elements of local energy density operators at four cells with positions $(x,y,z)=(1,1,0), (1,2,1), (0,1,0), (1,0,0)$ as functions of the eigenenergy for the $(n_x,n_y,n_z)=(2,3,3)$ hyperhoneycomb lattice Hamiltonian under fixed boundary conditions at $g=0.3$. The red line is the average of $\langle E_n|O|E_n\rangle$ within equal-width energy windows that correspond to a total of $N_{\rm bin}=80$ bins of width $1.58$, and the red band is the standard deviation within each bin. 
    }
    \label{fig:eth}
\end{figure}

Figure~\ref{fig:eth} shows the eigenstate expectation values of the energy densities $H_{xyz}$ at four lattice cells, $(x,y,z)=(1,1,0),(1,2,1),(0,1,0),(1,0,0)$, as functions of the eigenenergy $E_n$. In all four cases, the eigenstate expectation values (blue points) are distributed over a narrow band around a smooth function of energy, which is consistent with the ETH ansatz~\eqref{eqn:eth}. The solid red line is obtained by taking averages within equal-width energy windows, each of width $1.58$. In total, there are $N_{\rm bin}=80$ such windows. The red band denotes the standard deviation of the finite-size samples within each window. Demonstrating the exponential decay in the fluctuations as the lattice size increases calls for diagonalizing larger lattice Hamiltonians, which is beyond current classical computing capabilities.

\begin{figure}[h]
    \centering
    \includegraphics[width=1\linewidth]{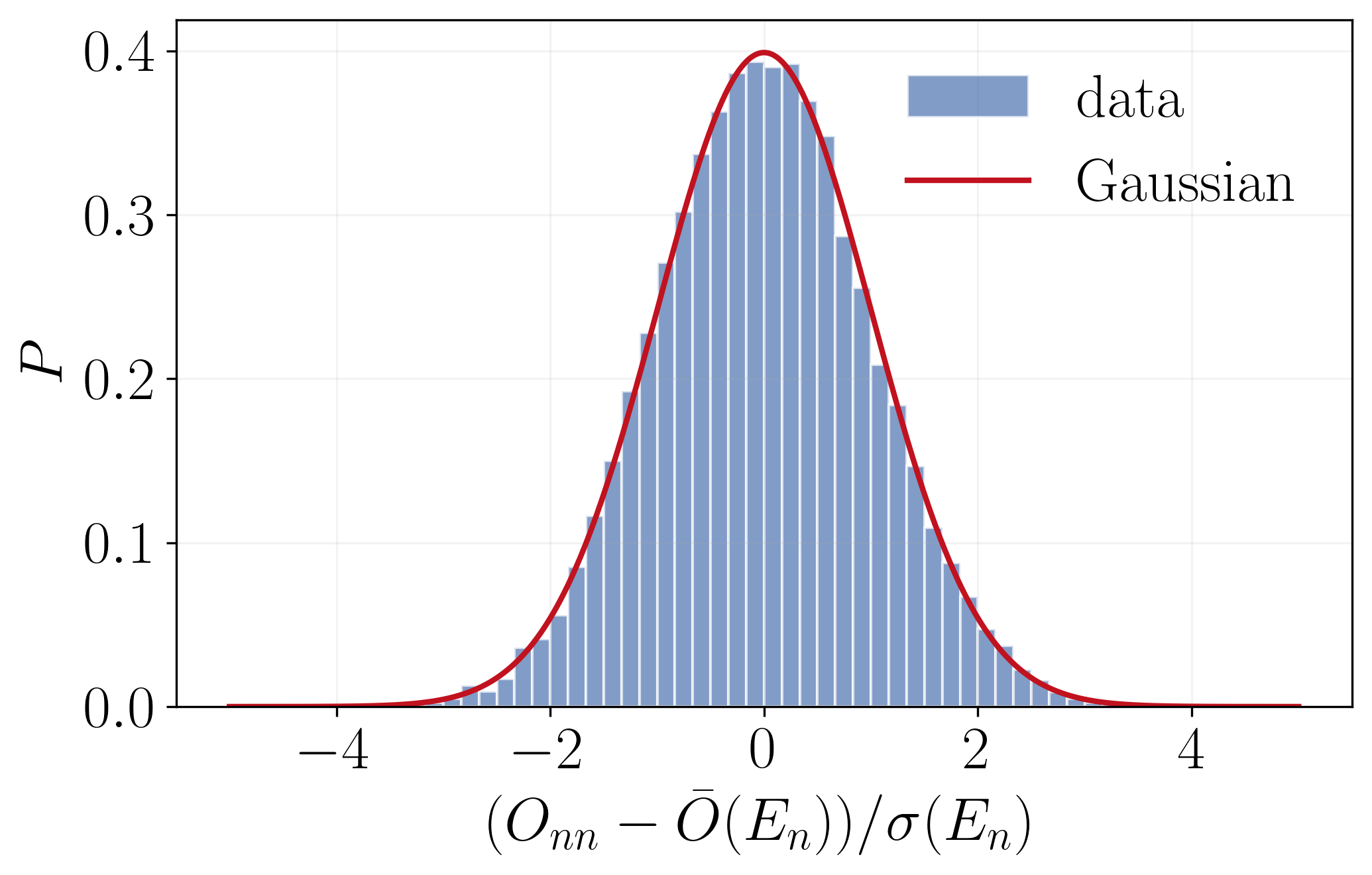}
    \caption{Distribution of normalized fluctuations $[O_{nn}-\bar{O}(E_n)]/\sigma(E_n)$ in the diagonal matrix elements of the local energy density operator at cell 121 from the central $20\%$ of the eigenstates in the spectrum. The corresponding energy range is $[2.63,9.09]$. The red line is the standard Gaussian distribution with zero mean and unit variance. The bin size of the histogram is $0.132$.
    }
    \label{fig:eth_gaussian}
\end{figure}

We also investigate the distribution of the fluctuations of the diagonal matrix elements around the microcanonical ensemble average. We take the diagonal matrix elements of $H_{121}$ from the central $20\%$ of the eigenstates, which correspond to the energy range $[2.63,9.09]$. We compute the fluctuation of each matrix element as $\delta O_n=O_{nn}-\bar{O}(E_n)$ and normalize it as $\delta O_n/\sigma(E_n)$, where $O_{nn}$ is an abbreviation for $\langle E_n| O |E_n\rangle$, and $\bar{O}(E_n)$ and $\sigma(E_n)$ are the average and standard deviation of the diagonal matrix elements from the 200 neighboring eigenstates of the state $|E_n\rangle$. The distribution of $\delta O_n/\sigma(E_n)$ is shown in Fig.~\ref{fig:eth_gaussian} as a histogram with bin size $0.132$. It agrees reasonably well with the standard Gaussian distribution with zero mean and unit variance (solid red line), confirming the ETH ansatz.

\begin{figure}[t]
    \centering
    \includegraphics[width=1\linewidth]{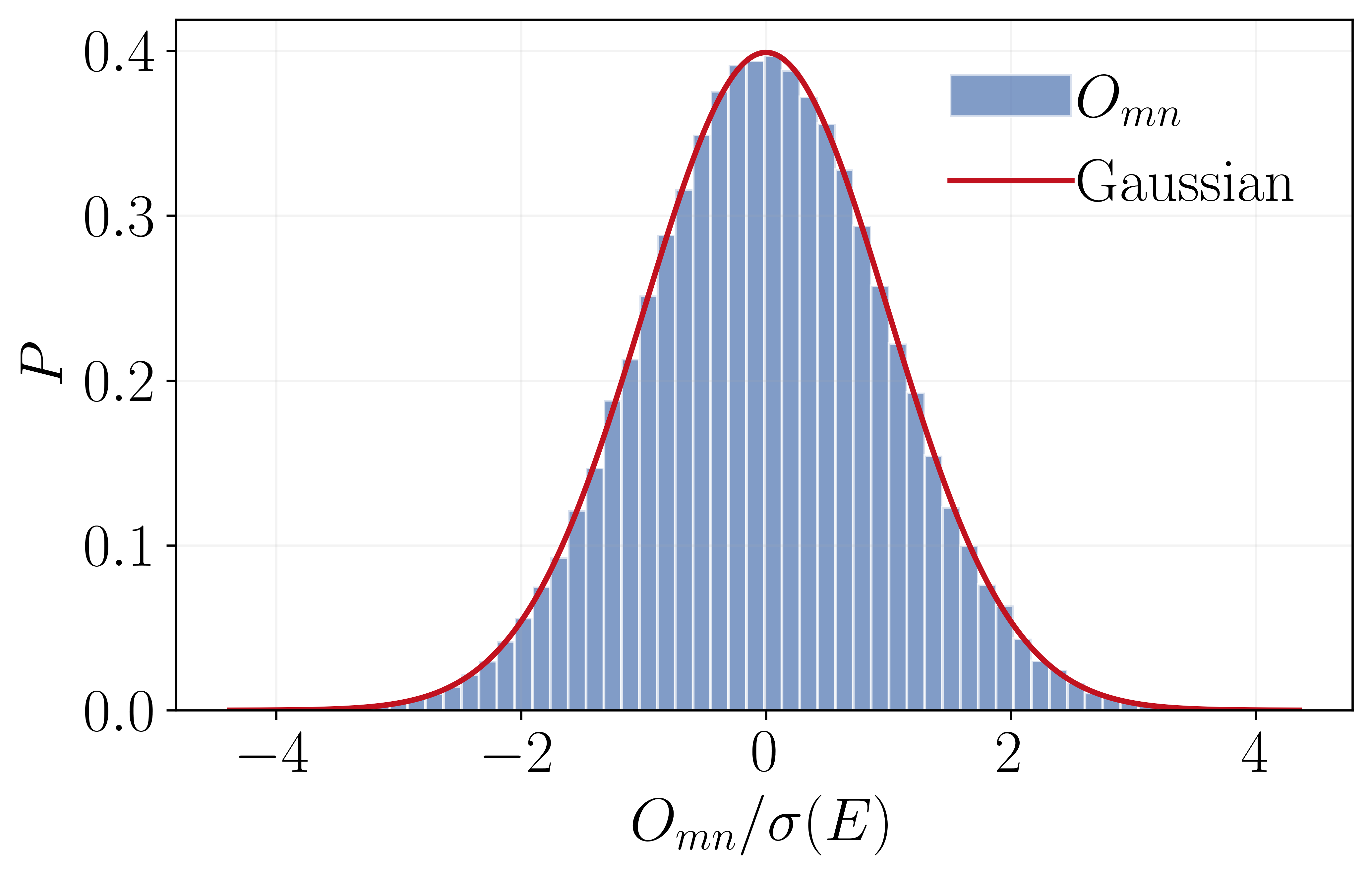}
    \caption{Distribution of normalized fluctuations $O_{mn}/\sigma(E)$ in the off-diagonal matrix elements of the local energy density operator at cell 121 from the energy window $[5.7,6]$ that contains $615$ eigenstates. We only take the off-diagonal matrix elements with $m<n$ since the operator is Hermitian, giving $188805$ elements in total. The red line is the standard Gaussian distribution with zero mean and unit variance. The bin size of the histogram is $0.1456$.
    }
    \label{fig:offdiag_gaussian}
\end{figure}

Moreover, we analyze the off-diagonal matrix elements of the same energy density operator $H_{121}$, which are purely fluctuations. We choose all $615$ eigenstates in the energy range $[5.7,6]$ in the spectrum and compute the off-diagonal matrix elements $\langle E_m | H_{121}|E_n\rangle$ for $m<n$. We normalize these matrix elements by the standard deviation $\sigma(E)$ of the sample, where the argument $E$ can be understood as representing the small energy window $[5.7,6]$. Figure~\ref{fig:offdiag_gaussian} depicts the distribution of these normalized off-diagonal matrix elements with a bin size of 0.1456, which agrees well with the standard Gaussian distribution shown as the solid red line.

The standard deviations $\sigma(E_n)$ and $\sigma(E)$ extracted from the data samples used in Figs.~\ref{fig:eth_gaussian} and~\ref{fig:offdiag_gaussian} correspond physically to $e^{-S(\bar{E})/2}f_{H_{121}}(\bar{E},\omega)$ for $\omega=0$ and $0<\omega\leq 0.3$, respectively. By using the density of states, one can further obtain the $f_{H_{121}}(\bar{E},\omega)$ function, which encodes transport properties of the lattice system. We do not pursue this here and leave it to future studies on simulating non-equilibrium dynamics and energy-momentum transport.

\subsection{Entanglement Entropy}
We have tested the ETH ansatz for the local energy density operators. The ETH is expected to hold for all local operators on a given spatial support. This can be summarized as the subsystem ETH~\cite{Dymarsky:2016ntg}, which states that a small subsystem in an excited eigenstate is thermal, 
\begin{equation}
\rho_A=\mathrm{Tr}_{B}\ket{E_n}\bra{E_n}\simeq \rho_{A}(\beta)\,, 
\end{equation}
where $A$ denotes a subset of qubits with complement $B$, $\rho_A(\beta)=e^{-\beta H_A}/Z_A$, $Z_A=\mathrm{Tr}_A(e^{-\beta H_A})$, and the temperature is fixed by requiring the thermal energy on the whole lattice equal the energy of the eigenstate, i.e., $\mathrm{Tr}[\rho(\beta)H]=E_n$ with $\rho(\beta)=e^{-\beta H}/\mathrm{Tr}(e^{-\beta H})$. The Hilbert space of our minimally truncated lattice theory factorizes since we solve the Gauss's law constraints everywhere and the qubit states represent only gauge singlet states.

We thus expect the entanglement entropy of an excited eigenstate to behave like the thermal entropy, as long as the subsystem size is less than half of the total size. Specifically, we expect the volume law
\begin{align}
    S_A(V_A,\beta) = -{\rm Tr}(\rho_A \ln \rho_A) \simeq s(\beta) V_A \,,
\end{align}
where $V_A$ is the volume of the subsystem $A$ and $s(\beta)$ is the thermodynamic entropy density. In particular, at infinite temperature $\beta=1/T = 0$, the corresponding eigenstates lie near the middle of the spectrum, with energies close to
\begin{equation}
\begin{aligned}
    E_\infty = \lim_{\beta \rightarrow0}\mathrm{Tr}\left(\frac{e^{-\beta H}}{Z}H\right) = \mathrm{Tr}\left(\frac{1}{d}H\right)=\frac{1}{d}\sum_nE_n \,,
\end{aligned}
\end{equation}
where $d$ is the dimension of the total Hilbert space. These eigenstates behave like Haar random states. The average entanglement entropy of a Haar random state is given by the Page formula~\cite{Page:1993df}:
\begin{equation}
    S_{\rm Page}(N_A)=\sum_{k=d_B+1}^{d_Ad_B}\frac{1}{k}-\frac{d_A-1}{2d_B},\ (d_A\le d_B)\,,
\end{equation}
where $d_A$ and $d_B$ denote the Hilbert space dimensions of the subsystem $A$ and its complement $B$, respectively.

On the other hand, for the ground state and low-lying eigenstates, the entanglement entropy is expected to follow the area law in gapped systems
\begin{align}
    S_A(V_A,\beta=\infty) \propto V_A^{2/3} \,.
\end{align}

\begin{figure}[ht]
    \centering
    \includegraphics[width=1\linewidth]{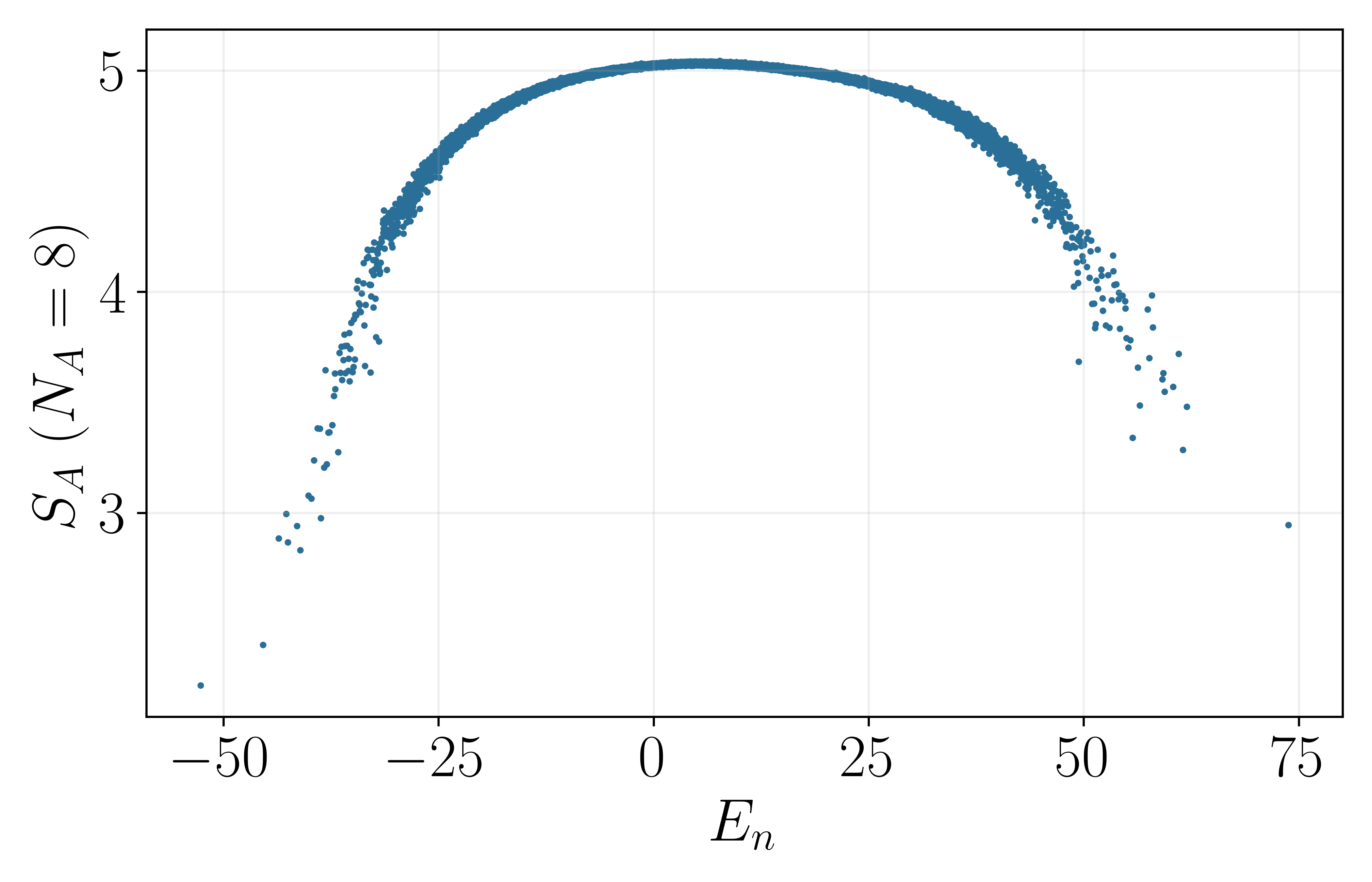}
    \caption{Half-system entanglement entropy $S_A(N_A=8)$ as a function of the excitation energy $E_n$ for the $(n_x,n_y,n_z)=(2,3,3)$ hyperhoneycomb lattice Hamiltonian under fixed boundary conditions at $g=0.3$.
    }
    \label{fig:eeHalfLA}
\end{figure}

Figure~\ref{fig:eeHalfLA} shows the half-system entanglement entropy $S_A(N_A=N_q/2)$ as a function of $E_n$ for the $(n_x,n_y,n_z)=(2,3,3)$ hyperhoneycomb lattice Hamiltonian under fixed boundary conditions at $g=0.3$, corresponding to $N_q=16$ qubits. The entanglement entropy first increases and then decreases as $E_n$ increases, reaching its maximum in the middle of the spectrum that corresponds to infinite temperature. We observe no highly excited states with low entanglement in the middle of the spectrum, implying that the minimally truncated lattice theory has no scar states at $g=0.3$. We expect this to be the case for all ergodic couplings.

\begin{figure}[ht]
    \centering
    \includegraphics[width=1\linewidth]{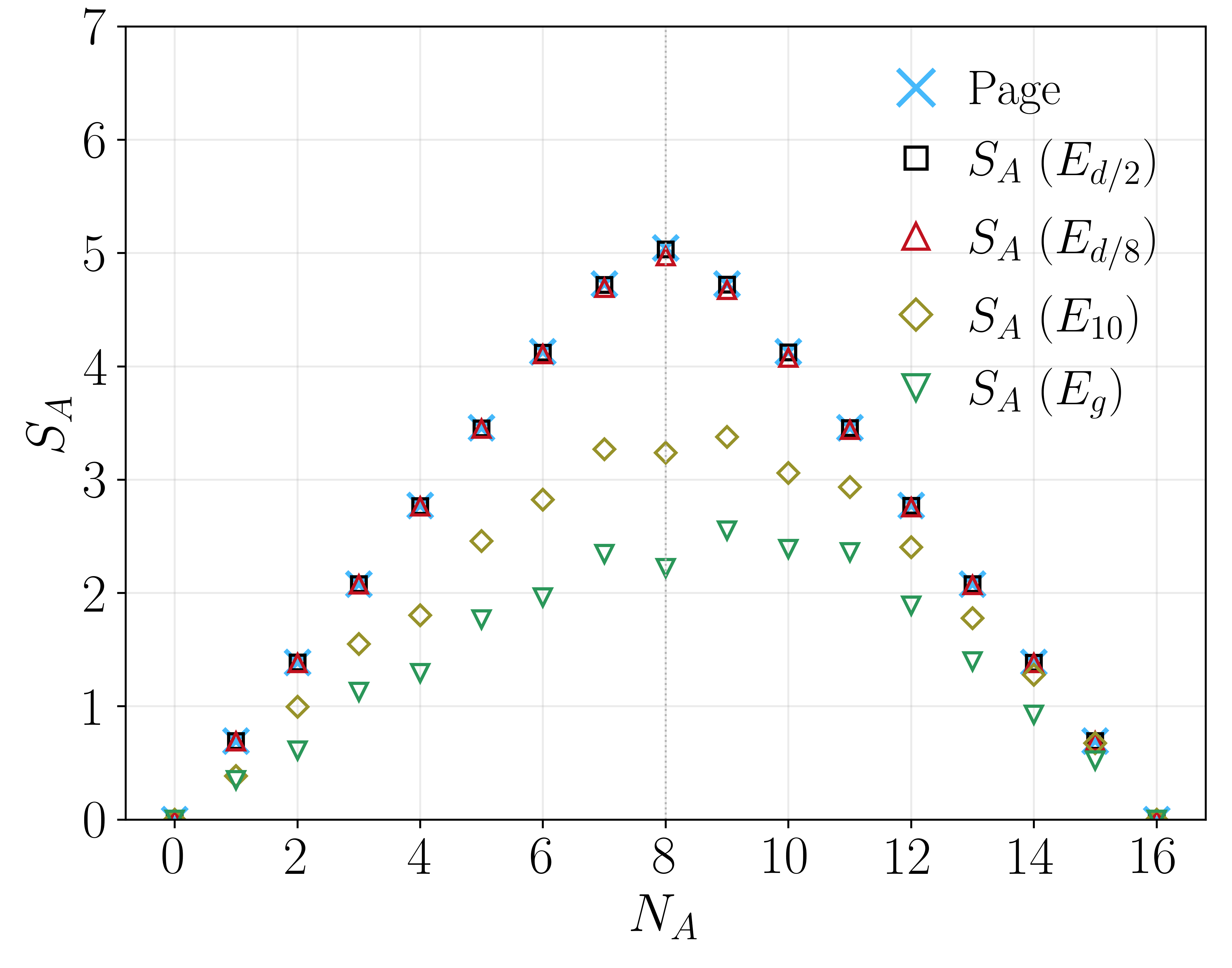}
    \caption{Entanglement entropies $S_A$ of different eigenstates as functions of the subsystem size (number of qubits) $N_A$ for the $(n_x,n_y,n_z)=(2,3,3)$ hyperhoneycomb lattice Hamiltonian under fixed boundary conditions at $g=0.3$. The subsystem $A$ consists of the first $N_A$ qubits, which are ordered first by their unit-cell coordinate $(x,y,z)$ in lexicographic order, $(0,0,0),(0,0,1),(0,0,2),(0,1,0),\dots,(1,2,2)$, and then, within each unit cell, by qubit type in the order $a,b,c,\beta$. The blue cross markers show the Page curve values for Haar random states. The hollow symbols show the entanglement entropies of different eigenstates, where the subscript represents the position of the eigenstate in the spectrum: $E_g$ is the ground state, $E_{10}$ is the 10th lowest eigenstate, $E_{d/8}$ and $E_{d/2}$ are the $(d/8)$th and $(d/2)$th eigenstates, with $d=2^{16}$ being the dimension of the Hilbert space.
    }
    \label{fig:page}
\end{figure}

Figure~\ref{fig:page} shows $S_A$ as a function of the number of qubits $N_A$ in the subsystem for a few eigenstates with drastically different energies: the ground state energy $E_g$, the 10th lowest eigenenergy $E_{10}$, the eigenenergy $E_{d/8}$ one eighth of the way up the spectrum from the ground state, and the one at the very center of the spectrum $E_{d/2}$. The subsystem $A$ consists of the first $N_A$ qubits, which are ordered first by their unit-cell coordinate $(x,y,z)$ in lexicographic order, $(0,0,0),(0,0,1),(0,0,2),(0,1,0),\dots,(1,2,2)$, and then, within each unit cell, by qubit type in the order $a,b,c,\beta$. 
At very high excitation energies such as $E_{d/8}$ and $E_{d/2}$, the entanglement entropy grows linearly for $N_A$ less than half the total number of qubits, consistent with the volume law. It can also be seen that for these eigenstates near the middle of the spectrum, the entanglement entropy closely follows the Page curve for Haar random states. The maximal entanglement stored in these states is a natural consequence of the ETH. Eigenstates that are much further away from the middle of the spectrum have lower entanglement entropies and deviate from the volume law. The entanglement entropy of the ground state is expected to follow the area law. Because the subsystem $A$ is defined by the first $N_A$ qubits in the ordering described above instead of a spatial region, $N_A$ is not proportional to the spatial volume of the subsystem, and thus the entropy is not expected to be strictly proportional to $N_A^{2/3}$. This is also the reason why the $N_A$ dependence of $S_A$ is not very smooth and is not symmetric with respect to the half system size $N_A=8$. Nevertheless, the half-system entanglement entropy $S_A(N_A=8)$ of the ground state is roughly half of the maximal value, which is consistent with the area-law estimate $N_A^{2/3}/N_A = 1/2$ at $N_A=8$. At excitation energies away from the middle of the spectrum, eigenstates deviate from the maximal entanglement limit, but their entanglement entropies are still expected to follow the predictions of the subsystem ETH, see e.g., Fig.~5 in Ref.~\cite{Ebner:2024qtu}.

\section{Conclusions}
\label{sec:conclusions}

In this work, we showed that the spatial-gauge-singlet states in the Kogut-Susskind Hamiltonian of the minimally truncated $3+1$D SU(2) lattice gauge theory on the hyperhoneycomb lattice can be mapped onto qubits faithfully by imposing Gauss's law at each vertex and eliminating the gauge redundancy. The number of qubits required for an $n \times n \times n$ lattice is $N_q=2n^3+O(n^2)$, i.e., two qubits per unit cell with a subleading surface correction. We explicitly wrote the Hamiltonian for the singlet states in terms of Pauli strings for an arbitrary lattice under fixed boundary conditions.

We also showed that this model can thermalize in the sense of eigenstate thermalization by demonstrating consistency with the ETH prediction on a $2\times 3\times 3$ lattice under fixed boundary conditions. After exactly diagonalizing the lattice Hamiltonian that describes only singlet states, we identified the coupling range of quantum ergodicity using level statistics. For the ergodic coupling $g=0.3$, we showed that the diagonal matrix elements of local energy density operators are smooth functions of the eigenenergy, with approximately Gaussian fluctuations in both the diagonal and off-diagonal matrix elements. Finally, we computed the entanglement entropies for all the eigenstates and found no scar states that have high excitation energy but low entanglement. The entanglement entropies of highly excited eigenstates obey the volume law and approach the Page curve for Haar random states. On the other hand, the entanglement entropy of the ground state follows the area law.

Because the minimally truncated SU(2) Hamiltonian on the hyperhoneycomb lattice is already written in terms of Pauli matrices, real-time simulation can be directly implemented on quantum computers via Trotterization, which we plan to pursue in the future. This would enable us to study larger lattice systems and simulate thermalization in non-equilibrium dynamics. Moreover, future work could extend this Hamiltonian by increasing the truncation $j_{\rm max}$, including fermions, and generalizing the setup to the SU(3) case. In the long run, it would be interesting to compare the Kogut-Susskind Hamiltonian on the hyperhoneycomb lattice in the electric basis with other lattice Hamiltonian approaches.

\begin{acknowledgments}
This work is supported by the U.S. Department of Energy, Office of Science, Office of Nuclear Physics, IQuS under Award Number DOE (NP) Award DE-SC0020970 via the program on Quantum Horizons: QIS Research and Innovation for Nuclear Science. This work was enabled, in part, by the use of advanced computational, storage and networking infrastructure provided by the Hyak supercomputer system at the University of Washington. 
\end{acknowledgments}

\bibliography{main.bib}
\end{document}